\documentclass[preprint,showpacs,preprintnumbers,amsmath,amssymb]{revtex4}

\usepackage{graphicx}% Include figure files
\usepackage{dcolumn}% Align table columns on decimal point
\usepackage{bm}% bold math
\usepackage{color}

\def\gsim{\mathop {\vtop {\ialign {##\crcr 
$\hfil \displaystyle {>}\hfil $\crcr \noalign {\kern1pt \nointerlineskip } 
$\,\sim$ \crcr \noalign {\kern1pt}}}}\limits}
\def\lsim{\mathop {\vtop {\ialign {##\crcr 
$\hfil \displaystyle {<}\hfil $\crcr \noalign {\kern1pt \nointerlineskip } 
$\,\,\sim$ \crcr \noalign {\kern1pt}}}}\limits}

\begin{document}

%\preprint{APS/123-QED^{*}}

\title{
On anomalous temperature dependence of relaxation rate measured by $\mu$SR in 
$\alpha$-YbAl$_{0.986}$Fe$_{0.014}$B$_4$
}

\author{Kazumasa Miyake}
\affiliation{
Center for Advanced High Magnetic Field Science, Osaka University, Toyonaka, Osaka 560-0043, Japan
}%Lines break automatically or can be forced with \\
\email{miyake@mp.es.osaka-u.ac.jp}
%\affiliation{%
%Authors' institution and/or address\\
%This line break forced with \textbackslash\textbackslash
%}%

\author{Shinji Watanabe}
% \homepage{http://www.Second.institution.edu/~Charlie.Author}
\affiliation{
Department of Basic Sciences, Kyushu Institute of Technology, Kitakyushu, {Fukuoka} 
804-8550, Japan
%Second institution and/or address\\
%This line break forced% with \\
}%

\date{11 June 2017; Revised 25 July 2018; Published 14 August 2018}
% It is always \today, today,
             %  but any date may be explicitly specified

\begin{abstract}
Recently, it was reported by MacLaughlin {\it et al}. in Phys. Rev. B {\bf 93}, 214421 (2016) 
that $\alpha$-YbAl$_{0.986}$Fe$_{0.014}$B$_4$ exhibits an anomalous temperature dependence 
in the relaxation rate $1/T_{1}$ of $\mu$SR, and stressed that such temperature dependence cannot be 
understood by the scenario based on the quantum critical valence transition (QCVT) while this compound 
exhibits a series of the non-Fermi liquid behaviors explained by the theory of the QCVT. 
In this paper, we point out that the anomalous temperature dependence in $1/T_{1}$ can 
be understood semi-quantitatively 
{
by assuming that the attraction of a screening cloud of conduction electrons about the
$\mu^{+}$ induces a local magnetic moment}
%by taking account of the effect of the $\mu^{+}$ that extracts 
%conduction electrons around it, and in turn induces the local magnetic moment 
arising from a 4f hole on the Yb ion, giving rise to the Kondo effect between heavy quasiparticles. 
\end{abstract}

\pacs{Valid PACS appear here}% PACS, the Physics and Astronomy
                             % Classification Scheme.
%\keywords{Suggested keywords}%Use showkeys class option if keyword
                              %display desired
\maketitle

%\section{\label{sec:level1}First-level heading:
\section{Introduction}
$\beta$-YbAlB$_4$ exhibits unconventional non-Fermi liquid  properties in the low temperature ($T$) 
region $T<10\,$K {not only} at ambient condition \cite{Nakatsuji} 
{but under a range of pressures \cite{Tomita}}. 
The critical exponents of a series of physical quantities were shown to 
follow those given by a theory of quantum critical valence transition (QCVT)  
\cite{Watanabe1,Watanabe2, Miyake} 
which also explains the unconventional non-Fermi liquid  properties of other systems exhibiting 
the same critical exponents that were observed in  
YbCu$_{5-x}$Al$_x$ ($x=3.5$) \cite{Bauer,Seuring}, 
YbRh$_2$Si$_2$ \cite{Trovarelli,Ishida}, and quasi-crystal compound 
Yb$_{15}$Al$_{34}$Au$_{51}$ \cite{Deguchi} in a range of pressures including ambient pressure and 
quasi-crystal-approximant Yb$_{14}$Al$_{35}$Au$_{51}$ under pressure 
$P\simeq 1.8\,$GPa \cite{Matsukawa}. 
{Recently, the $T/B$ scaling behavior observed in $\beta$-YbAlB$_4$} 
{\cite{Matsumoto2} and  Yb$_{14}$Al$_{35}$Au$_{51}$ \cite{Matsukawa}} have   
{also been shown to be explained 
from the theory of the QCVT \cite{Watanabe3}.}
On the other hand, a sister compound  $\alpha$-YbAlB$_4$ 
follows properties of {the conventional heavy-electron metal with an intermediate valence of 
Yb} \cite{Matsumoto}. 
However, $\alpha$-YbAl$_{1-x}$Fe$_x$B$_4$ ($x=0.014$) exhibits the same criticality as 
$\beta$-YbAlB$_4$ { and the drastic change in valence of Yb \cite{Kuga}.}  
This strongly suggests that the scenario of QCVT is valid. Nevertheless, 
it was recently reported in Ref.\ \cite{MacLaughlin} that $\alpha$-YbAl$_{0.986}$Fe$_{0.014}$B$_4$ 
exhibits an anomalous $T$ dependence in the relaxation rate $1/T_{1}$ measured by $\mu$SR 
(muon spin rotation),  which cannot be {simply} understood by the scenario based on the QCVT. 

\begin{figure}[h]
\begin{center}
\rotatebox{0}{\includegraphics[width=0.7\linewidth]{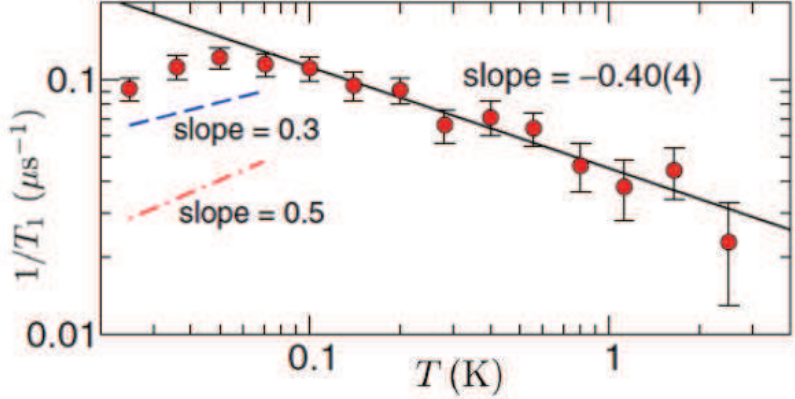}}
\caption{(color online) 
Temperature dependence (in the common logarithmic scale) of the relaxation rate $1/T_{1}$ 
 (in the common logarithmic scale) measured by  $\mu$SR experiment \cite{MacLaughlin}.  
}
\label{Relax_Rate_EXP}
\end{center}
\end{figure}

Figure \ref{Relax_Rate_EXP} shows the $T$ dependence in the muon relaxation rate $1/T_{1}$ of 
$\alpha$-YbAl$_{1-x}$Fe$_x$B$_4$ ($x=0.014$) at ambient pressure \cite{MacLaughlin}. 
The behavior at $T<0.05\,$K is consistent with the prediction by the theory of QCVT , i.e., 
$1/T_{1}T\propto T^{-\zeta}$ with weakly $T$-dependent exponent 
$\zeta$ ($0.5<\zeta<0.7$) \cite{Watanabe1,Watanabe2, Miyake}. On the other hand, 
the behavior $1/T_{1}T\propto T^{-1.4}$ (at $T>0.1$K),  is entirely different, which 
{seems to have} led 
the authors of Ref.\ \cite{MacLaughlin} to 
skepticism toward the QCVT {scenario}. The purpose of this paper is to give a possible  
explanation to this puzzling behavior by taking account of an influence of $\mu^{+}$ on the electronic 
state around it.  Namely, $\mu^{+}$ attracts conduction electrons which 
{lowers the crystalline electric field (CEF) levels of 4f holes for the Yb ions} 
{neighboring} it, inducing the local magnetic moment which should give rise to 
the Kondo effect that enhances the relaxation rate $1/T_{1}$ of $\mu^{+}$ through the 
conduction electrons around the Yb ion. 

%%%%%%%%%%%%%%%%%%%%%%%%%%%%%%%%%%%%%%%%%%%%%%%%%%%%%%%%%%%%%%%%%%%%%%%%%%%%%%%
\section{Physical Picture}
\begin{figure}[h]
\begin{center}
\rotatebox{0}{\includegraphics[width=0.9\linewidth]{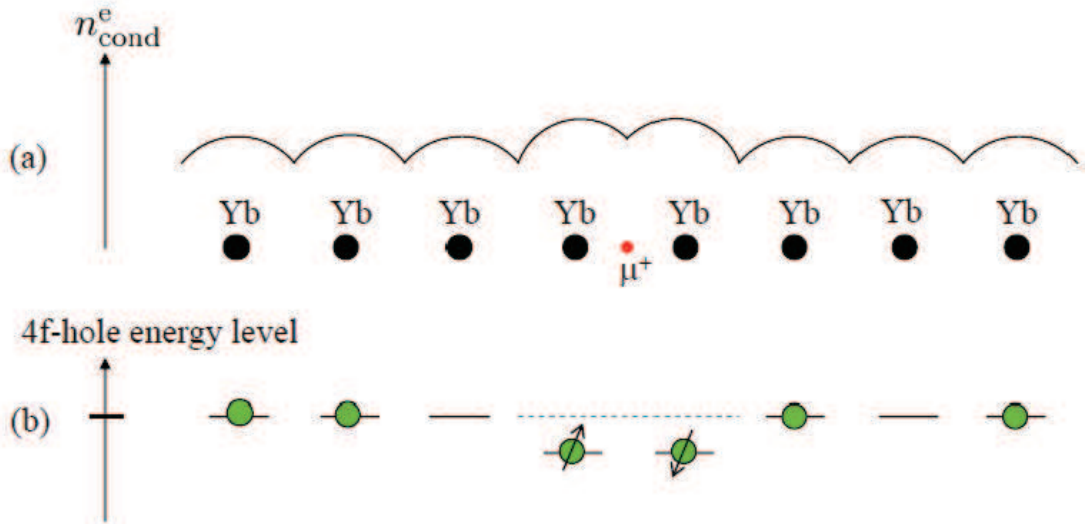}}
\caption{(color online)
(a) Schematic picture on electron distribution of conduction electrons attracted by $\mu^{+}$.  
(b) Snapshot of the { lowest CEF} energy level of 4f hole at Yb site adjacent to $\mu^{+}$.   
}
\label{Electron_Distribution}
\end{center}
\end{figure}

Figure\ \ref{Electron_Distribution}(a) shows a schematic picture of distribution of conduction electrons 
modified by the existence of $\mu^{+}$. Namely, the positive charge of $\mu^{+}$ attracts conduction 
electrons so that the density of conduction electrons should increase {neighboring} $\mu^{+}$. 
Figure\ \ref{Electron_Distribution}(b) shows a snapshot of the 
{lowest CEF energy levels}  
$\varepsilon_{\rm f}$s of 4f holes in Yb ions around  $\mu^{+}$, which are 
modified by the increase in conduction electrons density, and the distribution of 4f holes in the 
valence fluctuating situation. A crucial point is that the 4f-hole's energy level around $\mu^{+}$ 
is decreased by the repulsive Coulomb interaction suffered from the excess conduction electrons,  
so that  there arises a localized spin of 4f holes at Yb sites adjacent to $\mu^{+}$. 
{
This effect can be rephrased on the hole picture for conduction electrons. Namely,  
$\mu^{+}$ decreases the density of conduction holes around it, making the energy level 
$\varepsilon_{\rm f}={\bar \varepsilon}_{\rm f}+U_{\rm fc}n_{\rm c}$ of f holes decrease there 
\cite{Watanabe1,Watanabe2,Miyake}, 
where $n_{\rm c}$ is the number of conduction electron holes at the Yb site 
and ${\bar \varepsilon}_{\rm f}$ is the level of f holes without the $\mu^{{+}}$. 
}

This induced local moment {of the 4f hole} at the 
Yb site would cause the impurity Kondo effect between quasiparticles (consisting of 4f-hole lattice 
and conduction electron band through { the} renormalized hybridization 
{$V^{*}$}) around there, giving excess 
spin fluctuations of quasiparticles which in turn should give an excess relaxation of the $\mu^{+}$ spin 
through the hyperfine coupling between them.  
{
The exact position where the $\mu^{+}$ stops in the {crystal} is not known, as there was no statement 
about this in Ref.\ \cite{MacLaughlin}. However, the result of the anomalous exponent for the temperature 
dependence in the relaxation rate will not be altered because the effect is not sensitive to a position of 
$\mu^{+}$ so long as it stops in the crystal.
}

%%%%%%%%%%%%%%%%%%%%%%%%%%%%%%%%%%%%%%%%%%%%%%%%%%%%%%%%%%%%%%%%%%%%%%%%%%%%%%%
\section{Formulation for Relaxation Rate}
\begin{figure}[h]
\begin{center}
\rotatebox{0}{\includegraphics[width=0.6\linewidth]{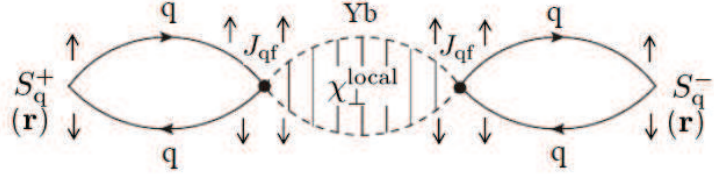}}
\caption{
Feynman diagram giving the relaxation rate $1/T_{1}T$ of the muon ($\mu^{+}$) through 
the hyperfine coupling with spin of quasiparticles at ${\bf r}$ (around the $\mu^{+}$) 
which is influenced by the local moment at Yb site induced by the effect of existence 
of $\mu^{+}$ particle itself.  
}
\label{Feynman_Diagram_1}
\end{center}
\end{figure}

For simplicity, hereafter, we treat the problem as the single impurity Kondo effect between the local moment 
at the Yb site (at the origin of space coordinate) and quasiparticles. 
The spin relaxation of $\mu^{+}$ stopped at ${\bf r}$ is given by the process of Feynman diagram 
shown in Fig.\ \ref{Feynman_Diagram_1} \cite{Moriya}, where $J_{\rm qf}$ is the bare exchange interaction 
between spins of quasiparticles and the localized 4f hole, and  $\chi_{\perp}^{\rm local}(\omega)$ is 
the dynamical transverse spin susceptibility of the localized 4f hole at the Yb site. 
{
Note that $J_{\rm qf}$ is proportional to the square of the renormalized hybridization $V^{*}$ 
between the localized 4f hole and the quasiparticles so that it is proportional to the 
mass renormalization amplitude $z$. Therefore, the dimensionless coupling constant 
$J_{\rm qf}N_{\rm F}^{*}$, 
$N_{\rm F}^{*}\propto z^{-1}$ being the renormalized density of states at the Fermi level of quasiparticles,  
is not subject to the effect of mass renormalization of the quasiparticles. 
}
{
The} {explicit form of} 
{$\chi_{\perp}^{\rm local}(\omega)$ is the retarded function of 
$\chi_{\perp}^{\rm local}({\rm i}\omega_{m})
\equiv \int_{0}^{1/T}{\rm d}\tau e^{{\rm i}\omega_{m}\tau}\langle S_{\rm f}^{+}(\tau)S_{\rm f}^{-}(0)\rangle$, 
where $S_{\rm f}^{\pm}$ are the spin-flip operators of the localized f electron and $\langle\cdots\rangle$ 
represents the 
thermal average.  }

\begin{figure}[h]
\begin{center}
\rotatebox{0}{\includegraphics[width=0.8\linewidth]{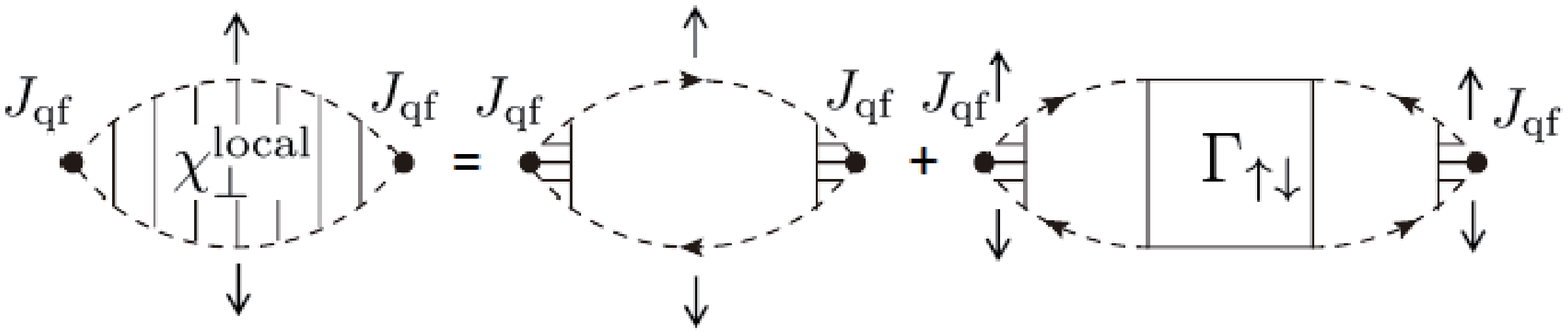}}
\caption{
Feynman diagram giving the renormalization of the local spin susceptibility $\chi_{\perp}^{\rm local}$ 
in Fig.\ \ref{Feynman_Diagram_1}. 
Triangles with a dot are the renormalized exchange interaction, a square 
$\Gamma_{\uparrow\downarrow}$ is the vertex correction causing the Kondo-Yosida singlet 
formation \cite{Yosida2,Yoshimori}, and dashed lines with arrow are the Matsubara Green function 
of pseudo fermion representing the localized 4f hole \cite{Abrikosov}. 
}
\label{Feynman_Diagram_2}
\end{center}
\end{figure}

The non-local dynamical transverse spin susceptibility $\chi_{\perp}^{\rm qp}({\bf r},\omega)$ 
of quasiparticles, appearing at both sides of $\chi_{\perp}^{\rm local}$ 
in Fig.\ \ref{Feynman_Diagram_1}, 
{is defined by the retarded function of 
\begin{equation}
\chi_{\perp}^{\rm qp}({\bf r},{\rm i}\omega_{m})\equiv
T\sum_{\epsilon_{n}}G^{\rm qp}_{\uparrow}({\bf r},{\rm i}\epsilon_{n}+{\rm i}\omega_{m})
G^{\rm qp}_{\downarrow}({\bf 0},{\rm i}\epsilon_{n}), 
\label{chi_local}
\end{equation}
where $G^{\rm qp}_{\sigma}({\bf r},{\rm i}\epsilon_{n})$ is the Matsubara Green function of 
the quasiparticles with the spin $\sigma$ ($=\uparrow$ or $\downarrow$). 
In the Fermi degenerate region ($T\ll D^{*}$, with $D^{*}$ being the 
effective Fermi energy} {or temperature} 
{of quasiparticles), the explicit form of  the retarded function 
(obtained by the analytic continuation ${\rm i}\omega_{m}\to\omega+{\rm i}\delta$) is given} by  
{
\begin{eqnarray}
\chi_{\perp}^{\rm qp}({\bf r},\omega)=N_{\rm F}^{*}\frac{2k_{\rm F}^{3}}{\pi}R(k_{\rm F}r)
+{\rm i}\pi\omega N_{\rm F}^{*2}\frac{[\sin(2k_{\rm F}r)]^{2}}{(k_{\rm F}r)^{2}},
\label{sus_cond}
\end{eqnarray} 
where $R(k_{\rm F}r)$ at $k_{\rm F}r\gsim 1$ represents the Friedel oscillations appearing in the 
Ruderman-Kittel-Kasuya-Yosida interaction \cite{Ruderman_Kittel,Kasuya,Yosida} and is defined as
\begin{eqnarray}
R(k_{\rm F}r)\equiv
-\frac{\cos(2k_{\rm F}r)}{(2k_{\rm F}r)^{3}}+\frac{\sin(2k_{\rm F}r)}{(2k_{\rm F}r)^{4}}, 
\label{RKKY}
\end{eqnarray}
while it approaches a dimensionless constant of the order of ${\cal O}(q_{\rm B}a)$,  
with $q_{\rm B}$ being the wavenumber of the Brillouin zone and $a$ being the lattice constant, 
in the limit $k_{\rm F}r\ll 1$.  }
%where $N_{\rm F}^{*}$ is the density of states at the Fermi level of quasiparticles. 
In deriving Eq.\ (\ref{sus_cond}), we have assumed the free dispersion for the quasiparticles band. 
Since the expression [Eq.\ (\ref{sus_cond})] is essentially $T$ independent, the crucial $T$ dependence 
arises from that of $\chi_{\perp}^{\rm local}(\omega)$ which is given by the Feynman diagram shown in 
Fig.\ \ref{Feynman_Diagram_2}. 

\begin{figure}[h]
\begin{center}
\rotatebox{0}{\includegraphics[width=0.7\linewidth]{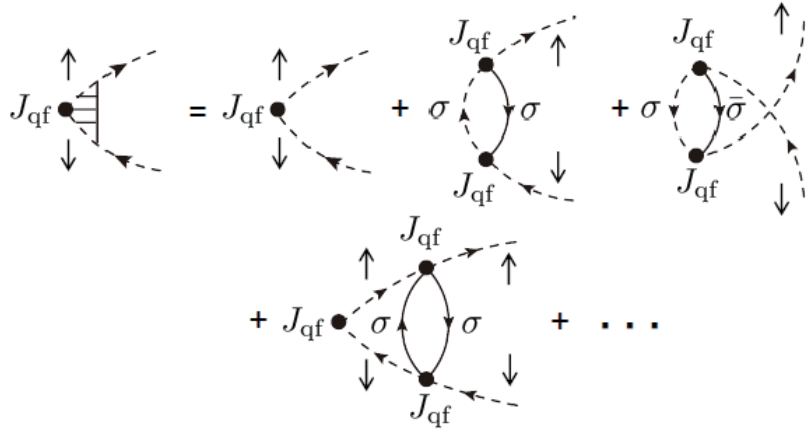}}
\caption{
Feynman diagram giving the renormalization of the exchange interaction $J_{\rm qf}$ 
corresponding to the spin-flip process shown 
in Figs.\ \ref{Feynman_Diagram_1} and \ref{Feynman_Diagram_2}. The label $\sigma$ and ${\bar \sigma}$ 
implies that the summation with respect to $\sigma$ ($\uparrow$ or $\downarrow$) and 
${\bar \sigma}$ ($\downarrow$ or $\uparrow$) is taken. 
}
\label{Exchange_Vertex}
\end{center}
\end{figure}

The triangle and square $\Gamma_{\uparrow\downarrow}$ in Fig.\ \ref{Feynman_Diagram_2} are 
renormalized exchange interaction and vertex correction expressing the effect of Kondo-Yosida 
singlet formation \cite{Yosida2,Yoshimori}, respectively.  The explicit form of the triangle is given by 
Fig.\ \ref{Exchange_Vertex} up to processes of the two-loop order,
{
and is known to increase by the Kondo renormalization effect \cite{PoormanScaling}, which is an 
origin of the anomalous relaxation rate. }

{
Concluding this section, we note that there also exists the relaxation process through the direct magnetic 
dipolar coupling between the localized spin of the 4f hole on the Yb ion and $\mu^{+}$ other than the process shown in Fig.\ \ref{Feynman_Diagram_1}.  However,}
{
this process gives only the $T$ independent contribution in $1/T_{1}$ reflecting the existence of the 
local moment of the 4f hole at the Yb site, 
so that it is masked by the anomalous contribution through 
the renormalization effect of $J_{\rm qf}$ shown in Fig.\ \ref{Exchange_Vertex} in the high temperature 
region $T \gsim T_{\rm K}$ as discussed in the next section. 
In the low temperature region $T\lsim T_{\rm K}$,} 
{it gives the  conventional temperature dependence in 
the relaxation rate $1/T_{1}$ as expected from the Korringa-Shiba relation at $T\ll T_{\rm K}$ \cite{Shiba},} 
{
which is also masked by the contribution given by the theory of QCVT \cite{Watanabe1}.  
}

%%%%%%%%%%%%%%%%%%%%%%%%%%%%%%%%%%%%%%%%%%%%%%%%%%%%%%%%%%%%%%%%%%%%%%%%%%%%%%%
\section{Anomalous Relaxation}
The temperature dependence of the relaxation rate $(1/T_{1}T)^{\rm local}_{1{\rm st}}$ 
arising from fluctuations 
of the local moment at Yb site, corresponding to the first term of the right-hand side (rhs) 
in Fig.\ \ref{Feynman_Diagram_2}, is given by 
\begin{eqnarray}
& &
\left(\frac{1}{T_{1}T}\right)^{\rm local}_{1{\rm st}}={{\tilde A}}_{\rm hf}^{{2}}
[J_{\rm qf}(T/D^{*})]^{2}
{
\left\{
\frac{{\rm Im}\chi_{0}^{\rm local}(\omega)}{\omega}\left[\chi_{\perp}^{\rm qp}({\bf r},0)\right]^{2}
\right.
}
\nonumber
\\
& &
{
\left.
\qquad\qquad\qquad\qquad\qquad\qquad\qquad
+2\frac{{\rm Im}\chi_{\perp}^{\rm qp}({\bf r},\omega)}{\omega}
\chi_{\perp}^{\rm qp}({\bf r},0)\chi_{0}^{\rm local}(0)
\right\},
}
\label{relaxation:1}
\end{eqnarray}
where ${{\tilde A}}_{\rm hf}$ is the {effective} hyperfine coupling constant 
between $\mu^{+}$ and quasiparticles, and 
$J_{\rm qf}(T/D^{*})$ is the renormalized exchange interaction given by the processes as 
shown in Fig.\ \ref{Exchange_Vertex}, and has the strong $T$ dependence characteristic to the 
Kondo effect {\cite{Alloul}}.  
{The effective hyperfine coupling ${\tilde A}_{\rm hf}$ 
is considered to arise mainly from the magnetic dipolar coupling between the muon and 
quasiparticles around it, while the direct hyperfine coupling between $\mu^{+}$ and electrons 
is quite small compared to that for usual nuclei used for NMR measurements.}

Here, we are adopting a scheme of the renormalization group (RG) approach in which the effects of 
intermediate states of quasiparticles with higher energies than the temperature $T$ are absorbed 
into the renormalized exchange interaction $J_{\rm qf}(T/D^{*})$ {\it \`a la} the {\it poorman}'s scaling 
approach {\cite{PoormanScaling,Tanikawa}}.  
The dynamical susceptibility $\chi_{0}^{\rm local}(\omega)$ 
of the local moment in Eq.\ (\ref{sus_cond}) represents that without 
the vertex correction and its imaginary part is proportional to $\omega/T$ in the limit $\omega \to 0$. 
Thus, 
\begin{equation}
\left(\frac{1}{T_{1}T}\right)^{\rm local}_{1{\rm st}}
\propto
\frac{[J_{\rm qf}(T/D^{*})]^{2}}{T}.
\label{relaxation:2}
\end{equation}

On the other hand, the contribution from the second term of the rhs in Fig.\ \ref{Feynman_Diagram_2} 
becomes important in the region $T\lsim T_{\rm K}^{*}$, with $T_{\rm K}^{*}$ being the Kondo temperature 
given by the one-loop order calculation \cite{PoormanScaling}, and 
works to suppress the relaxation rate through the 
Kondo-Yosida singlet formation \cite{Yosida2,Yoshimori}, leading to the Korringa-Shiba relation of 
the local Fermi liquid at $T\ll T_{\rm K}^{*}$ \cite{Shiba}. Therefore, in the region $T\lsim T_{\rm K}^{*}$, 
the relaxation rate $(1/T_{1}T)^{{\rm local}}$ due to 
the fluctuations of the local moment at the Yb site follows the relation 
\begin{equation}
\left(\frac{1}{T_{1}T}\right)^{{\rm local}}\propto {\rm const}. 
\label{relaxation:3}
\end{equation}
Namely, this contribution would be masked by that from the QCVT, 
$(1/T_{1}T)^{{\rm QCVT}}\propto T^{-\zeta}$ \cite{Watanabe1,Watanabe2, Miyake}, which dominates 
over the contribution $(1/T_{1}T)^{{\rm local}}$ [Eq.\ (\ref{relaxation:3})] in the low temperature 
region $T\lsim T_{\rm K}^{*}$. 

As a result, the relaxation rate $(1/T_{1}T)^{\rm local}_{1{\rm st}}$ 
[Eq.\ (\ref{relaxation:1})] dominates in the region $T\gsim T_{\rm K}^{*}$, 
and the exponent $\alpha$, giving an extra temperature dependence in the relaxation rate 
$(1/T_{1}T)^{\rm local}_{1{\rm st}}$, arises from that of the renormalized exchange interaction between 
quasiparticles and the localized 4f hole, $J_{\rm qf}(x)$,  
with $x$ being $x\equiv T/D^{*}$. Namely, $\alpha(x)$ in the region $T\gsim T_{\rm K}^{*}$ is defined by 
\begin{equation}
\left[J_{\rm qf}(x)\right]^{2}\equiv J_{0}^{2}\,x^{-\alpha(x)}, 
\label{alpha}
\end{equation}
where $J_{0}$ is the bare exchange interaction 
{ of $J_{\rm qf}$ at $T_{\rm K}^{*}\ll T\lsim D^{*}$} . 
Then, according to the relation Eq.\ (\ref{relaxation:1}), 
the relaxation rate $(1/T_{1}T)^{\rm local}_{1{\rm st}}$ is expressed as 
\begin{equation}
\left(\frac{1}{T_{1}T}\right)^{\rm local}_{1{\rm st}}\propto T^{\,-[1+\alpha(T/D^{*})]}.
\label{relaxation:4}
\end{equation}
With the use of Eq.\ (\ref{alpha}), the exponent $\alpha(x)$ is in turn given by  
\begin{equation}
\alpha(x)=-2\frac{\log[J_{\rm qf}(x)/J_{0}]}{\log\,x}. 
\label{alpha:2}
\end{equation}

%Other than the contribution $(1/T_{1}T)^{\rm local}_{1{\rm st}}$ [Eq.\ (\ref{relaxation:1})], there also exists 
%contributions from the product of the imaginary and real parts of $\chi_{\perp}^{\rm qp}({\bf r},\omega)$ 
%[Eq.\ (\ref{sus_cond})] and the real part of 
%$\chi_{\perp}^{\rm local}(\omega)$.   
%However, since the former term is essentially $T$ independent in 
%the region $T\ll D^{*}$, and the latter term gives at most only $1/T$ singularity at $T\gsim T_{\rm K}^{*}$ 
%and the local Fermi liquid behavior, $1/T_{1}T=$const., at $T\ll T_{\rm K}^{*}$, there arises no 
%contribution to the anomalous exponent $\alpha(x)$. 

%%%%%%%%%%%%%%%%%%%%%%%%%%%%%%%%%%%%%%%%%%%%%%%%%%%%%%%%%%%%%%%%%%%%%%%%%%%%%%%
\section{Temperature Dependence of Anomalous Exponent} 
\begin{figure}[h]
\begin{center}
\rotatebox{0}{\includegraphics[width=0.5\linewidth]{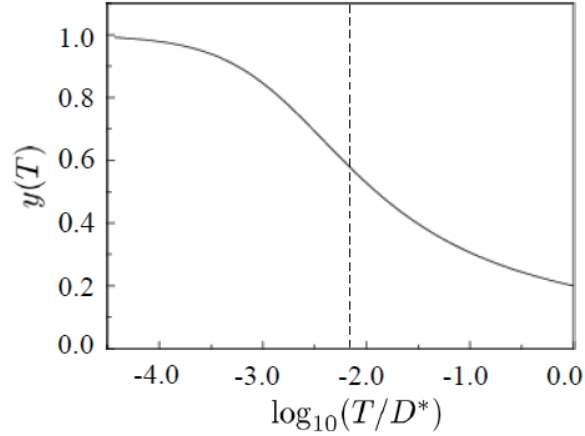}}
\caption{
Temperature dependence of the renormalized exchange interaction $y(T)$ as a function of 
$\log_{10}(T/D^{*})$ on the basis of the RG of the two-loop order for the bare interaction $y_{0}=0.2$. 
The dashed line corresponds to the Kondo temperature 
$T_{\rm K}^{*}\equiv D^{*}e^{-1/y_{0}}\simeq 0.67\times10^{-2}\,D^{*}$ given by the RG calculation of 
one-loop order. 
}
\label{Two_Loop_Scaling}
\end{center}
\end{figure}

The temperature dependence of the exponent $\alpha(T/D^{*})$ is derived by solving the 
two-loop RG evolution equation for the renormalized exchange interaction 
$J_{\rm qf}(x)$.  
 %For concise notation, hereafter,  we omit ``cf'' in $J_{\rm cf}$.  
The RG evolution equation for $y\equiv J_{\rm qf}N_{\rm F}^{*}$ on the two-loop order is given by 
\cite{Fowler_Zawadowski,Abrikosov_Migdal}
\begin{equation}
\frac{{\rm d}y}{{\rm d} t}=-y^{2}+y^{3},
\label{RG:1}
\end{equation}
where $t\equiv \log\,x$. 
This differential equation has a formal solution as 
\begin{equation}
\log\,\frac{y(y_{0}-1)}{y_{0}(y-1)}-\frac{1}{y}+\frac{1}{y_{0}}=-t,
\label{RG:2}
\end{equation}
where $y_{0}\equiv y(0)=J_{\rm qf}(0)N_{\rm F}^{*}$.  
The numerical relation between $y$ and $t$ [$=\log(T/D^{*})$] is easily obtained, 
as shown in Fig.\ \ref{Two_Loop_Scaling}, e.g., in the case of $y_{0}=0.2$. 

On the other hand, the RG evolution equation on the one-loop order is simplified and is given by  
\begin{equation}
\frac{{\rm d}y}{{\rm d} t}=-y^{2},
\label{RG:3}
\end{equation}
which was derived by Anderson on the idea of the poorman's scaling \cite{PoormanScaling}.   
The solution of Eq.\ (\ref{RG:3}) is explicitly given by 
\begin{equation}
y=\frac{y_{0}}{1+y_{0}t}=\frac{y_{0}}{1+y_{0}\log\,(T/D^{*})},
\label{RG:4}
\end{equation}
where the explicit $T$ dependence is shown in the second equality.  The Kondo temperature 
$T_{\rm K}^{*}$ is defined by the condition that the renormalized exchange interaction $y(T)$ diverges: 
i.e., $T_{\rm K}^{*}=D^{*}e^{-1/y_{0}}$ or $D^{*}\exp\,[-1/J_{\rm qf}(0)N_{\rm F}^{*}]$
{
although the divergence of  $y(T)$ at $T=T_{\rm K}$ is an artifact of insufficient approximation scheme. 
Nevertheless, it offers the characteristic temperature below which the Kondo-Yosida singlet state 
begins to be stabilized.  
}
Then, the exponent $\alpha(x)$ [Eq.\ (\ref{alpha:2})] is given by 
\begin{equation}
\alpha(x)=2\,\frac{\log\,[1+y_{0}\log\,x]}{\log\,x}. 
\label{alpha:3}
\end{equation}
Therefore, in the region $T\simeq D^{*}\gg T_{\rm K}^{*}$ (or $0<-\log\,x\ll 1$), 
the exponent $\alpha(x)$ becomes $T$ independent and is given by
\begin{equation}
\alpha(x)=2y_{0}=2J_{\rm qf}(0)N_{\rm F}^{*}. 
\label{alpha:4}
\end{equation}

\begin{figure}[h]
\begin{center}
\rotatebox{0}{\includegraphics[width=0.5\linewidth]{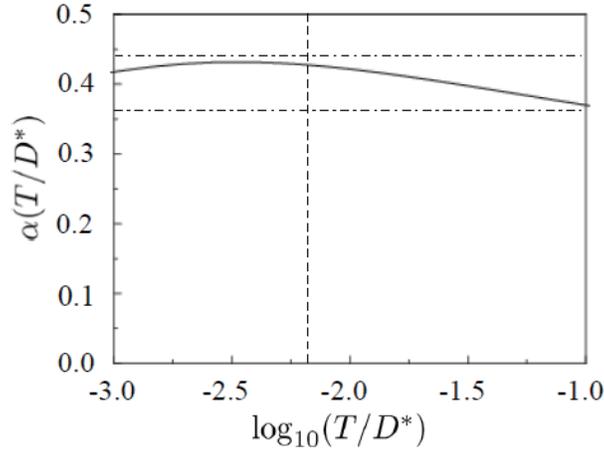}}
\caption{
Temperature dependence of the exponent $\alpha(T/D^{*})$ [Eq.\ (\ref{alpha:2})] as a function of 
$\log_{10}(T/D^{*})$ on the basis of the two-loop RG equation with the initial condition $y_{0}=0.2$. 
The dashed line corresponds to the Kondo temperature 
$T_{\rm K}^{*}\equiv D^{*}e^{-1/y_{0}}\simeq 0.67\times10^{-2}\,D^{*}$ given by the RG calculation of 
the one-loop order. 
The chain lines are lower and upper boundaries corresponding to the error bar in the experiment 
reported in Ref.\ \cite{MacLaughlin}. 
}
\label{Alpha_LogT}
\end{center}
\end{figure}

However, on the two-loop order, the exponent $\alpha(x)$ has a weak $T$ dependence. 
With the use of the numerical solution of Eq.\ (\ref{RG:2}) with an initial condition $y_{0}=0.2$, 
the temperature dependence in the exponent $\alpha(T/D^{*})$ [Eq.\ (\ref{alpha:2})] is given as 
Fig.\ \ref{Alpha_LogT}.  
It is remarked that the anomalous exponent $\alpha(T/D^{*})$ is almost 
$T$ independent in the region $T\gsim T_{\rm K}^{*}\simeq 0.67\times10^{-2}\,D^{*}$ 
and is located within the 
error bar of experiments reported in Ref.\ \cite{MacLaughlin}. This in turn implies that 
the bare coupling exchange interaction $J_{\rm qf}(0)$ between quasiparticles and localized 4f hole 
takes the value $J_{\rm qf}(0)N_{\rm F}^{*}=0.2$ which is { rather} 
difficult {a physical quantity 
to estimate} theoretically {\cite{WatanabeDMRG}}. 
Furthermore, the effective Fermi {temperature} $D^{*}$ of the quasiparticles 
should be $D^{*}\simeq 30\,$K considering 
that the Kondo temperature is estimated as $T_{\rm K}^{*}\simeq 0.2\,$K where the $T$ dependence of 
$1/T_{1}$ is expected to deviate from the scaling behavior $1/T_{1}\propto T^{-0.40\pm0.04}$, 
as shown in Fig.\ \ref{Relax_Rate_EXP}.  It should be mentioned, however, that $D^{*}\simeq 30\,$K 
is far smaller than the characteristic temperature $T^{*}=200\,$K estimated from 
the $T$ dependence of the Sommerfeld coefficient $C/T$ in Ref.\ \cite{Nakatsuji}.    

{
On the other hand, it turns out that the effective Fermi temperature $D^{*}\simeq 30\,$K is not a 
ridiculous possibility,  
if we compare the $T$ dependence of the specific heat of $\alpha$-YbAlB$_4$ with that of 
a typical heavy fermion system CeCu$_6$ 
\cite{Satoh}, in which $C/T$ begins to increase from $T_{\rm K}\simeq 3.5\,$K, which is identified 
with the effective Fermi temperature, and to reach in the low temperature  
$\lim_{T\to 0}\,C/T\simeq 1.6\times 10^{3}$ mJ/K$^2\cdot$mol(Ce).  
In the case of $\alpha$-YbAlB$_4$, $C/T$ begins to increase from ${{\tilde T}}\simeq 30\,$K and to reach 
$\lim_{T\to 0}\,C/T\simeq 1.3\times 10^{2}$ mJ/K$^2\cdot$mol(Yb) \cite{Matsumoto}.  
Therefore, if this ${\tilde T}\simeq 30\,$K is identified with the effective Fermi temperature  
$D^{*}\simeq 30\,$K, the behaviors of the specific heat in both systems are approximately 
related by changing the temperature scale by about ten times.  
It should be also remarked that CeCu$_6$ is located near the QCVT point which is approached 
under the pressure and the magnetic field \cite{Miyake,Raymond,Hirose, Miyake1} as in the 
case of $\alpha$-YbAl$_{1-x}$Fe$_x$B$_4$ ($x=0$) \cite{Matsumoto}. 
}

{W}ith these reservation{s}, the above result on the exponent $\alpha(T/D^{*})$ has 
{
offered a possible key concept to resolve} the puzzle on the anomalous $T$ dependence 
in the relaxation rate $1/T_{1}$ measured by $\mu$SR experiment \cite{MacLaughlin}.  

%%%%%%%%%%%%%%%%%%%%%%%%%%%%%%%%%%%%%%%%%%%%%%%%%%%%%%%%%%%%%%%%%%%%%%%%%%%%%%%
\section{Summary and Supplemental Discussions}
On the basis of the physical picture that the $\mu^{+}$ stopped at the interstitial in the crystal 
greatly influences the electronic state around it, it has been predicted that there arises the anomalous 
temperature dependence of the $\mu$SR relaxation rate $1/T_{1}$ observed in 
$\alpha$-YbAl$_{0.986}$Fe$_{0.014}$B$_{4}$ \cite{MacLaughlin}. Namely, the conduction electrons attracted 
by $\mu^{+}$ induce the local moment of the 4f hole on the Yb ion nearby and the Kondo effect is caused 
between the local moment and heavy quasiparticles, resulting in the excess contribution to the 
relaxation rate as $(1/T_{1})^{\rm local}\propto T^{-\alpha(T)}$ in the high temperature region 
$T\gsim T_{\rm K}^{*}$.  While the exponent $\alpha(T)$ depends on the exchange interaction 
$J_{\rm qf}$ between the quasiparticles and the local moment of the 4f hole and is weakly $T$ dependent, 
it is possible to choose a reasonable set of parameter{s}, 
$J_{\rm qf}N_{\rm F}^{*}=0.2$ and $D^{*}=30\,$K 
(corresponding to $T_{\rm K}^{*}=0.2\,$K), to reproduce the observed value $\alpha=0.40\pm0.04$, 
as shown in 
Fig.\ \ref{Alpha_LogT}.  On the other hand, in the low temperature region $T\lsim T_{\rm K}^{*}$, 
the local moment forms the Kondo-Yosida singlet state with quasiparticles so that the local Fermi liquid 
behavior is recovered, i.e., $(1/T_{1}T)^{\rm local}\propto {\rm const.}$ However, 
this contribution is buried by the contribution due to the QCVT, 
$(1/T_{1}T)^{\rm QCVT} \propto T^{-\zeta}$ ($0.5<\zeta<0.7$), which is really observed at 
$T<0.05\,$K$\ll T_{\rm K}^{*}=0.2\,$K \cite{MacLaughlin}. 

{
Although we have discussed the case of zero magnetic field in the present paper, 
the effect of magnetic field is considered to be also crucial for anomalies of the relaxation rate 
because the quantum criticality of valence transition is considerably  influenced by 
the magnetic field as discussed in Ref.\ \cite{Watanabe4}.   Indeed, the magnetic field dependence 
of the relaxation rate in $\alpha$-YbAl$_{0.986}$Fe$_{0.014}$B$_{4}$ has some structure at 
$H\sim 4\,$Oe \cite{MacLaughlin}, which might have 
some relevance to the magnetic field effect mentioned above.  However,  detailed analyses are 
left for future study. 

Finally, we have put aside the issue of the possibility of forming a muonium because it seems to be excluded 
in bulk metallic systems as discussed in Refs. \cite{Rogers, Mott}.  This is because the screening effect 
on the Coulomb attractive potential from $\mu^{+}$ works to inhibit the existence of the bound electronic 
state (i.e., muonium). 
}
{
Note that the screening length $\lambda_{\rm s}$ estimated by the Thomas-Fermi formula \cite{Ziman}, 
which is valid also in  the heavy fermion system because the charge susceptibility is 
essentially unrenormalized \cite{Varma}, is given by 
\begin{eqnarray}
\lambda_{\rm s}=\sqrt{\frac{E_{\rm F}}{6\pi ne^{2}}}
= \frac{\pi}{2}\left(\frac{3}{\pi}\right)^{1/6}\sqrt{\frac{n^{-1/3}}{a_{\rm B}}}\,a_{\rm B},
\label{screening} 
\end{eqnarray}  
where $E_{\rm F}$ is the Fermi energy of free electron and $n$ is the carrier number density 
and $a_{\rm B}$ is the Bohr radius. 
If we adopt $n^{-1/3}=4$\AA\ assuming that each Yb ion supplies one mobile electron 
\cite{Nakatsuji,Matsumoto}, the screening length is estimated as $\lambda_{\rm s}\simeq 2.3$\AA .  
Therefore, the screening is far from perfect at the Yb site  so that a finite fraction of conduction electrons 
can be accumulated at the Yb site giving rise to the local moment of the 4f hole at the Yb site 
because the system is 
at criticality of the valence transition of the Yb ion, which justifies a physical picture as shown 
in Fig.\ \ref{Electron_Distribution}.  
}

\begin{acknowledgments}
We are grateful to S. Nakatsuji
for informing us {of} the result of 
$\mu$SR measurement and urging us to give its interpretation from the viewpoint of the 
QCVT scenario.   
 We are also grateful to K. Kuga and Y. Matsumoto for informing us of  unpublished 
data and for careful reading of the manuscript.  One of us (K.M.) benefited greatly from communications 
with H. Mukuda on the NMR relaxation rate of the host ions near the Kondo impurity{, 
and those with J. Quintanilla and H. Matsuura on the problem of muonium formation in metals.}  
This work was supported by a Grant-in-Aid for Scientific Research (Grants No. 24540378, No.  
{15K03542, No. 18H04326,} and No. 17K05555) from the Japan Society for the Promotion of Science. 
One of us (S.W.) was supported by JASRI 
(Grant No. 0046 in 2012B, 2013A, 2013B, 2014A, 2014B, and 2015A). 
\end{acknowledgments}

\newpage %Just because of unusual number of tables stacked at end
%%%%%%%%%%%%%%%%%%%%%%%%%%%%%%%%%%%%%%%%%%%%%%%%%%%%%%%%%%%%%%%%%%%%%%%%%%%%%%%
% References

\end{document}